\documentclass[aps,prl,showpacs,twocolumn]{revtex4}

\usepackage{amsmath}
\usepackage[dvips]{graphicx}
\usepackage{latexsym}
\usepackage{amsfonts}
\usepackage{amssymb}

\newcommand{\ket}[1]{\ensuremath{\left| #1 \right>}}

\begin{document}

\title{Experimental demonstration of bosonic commutation relation via superpositions of quantum operations on thermal light fields}

\author{A. Zavatta$^{1,2}$, V. Parigi$^{1,2}$, M. S. Kim$^3$, H. Jeong$^4$, and M. Bellini$^{1,2}$}

\affiliation{$^1$Istituto Nazionale di Ottica Applicata (INOA-CNR), L.go E. Fermi 6, 50125 Florence, Italy\\
$^2$LENS and Department of Physics, University of Firenze, 50019 Sesto Fiorentino, Florence, Italy\\
$^3$School of Mathematics and Physics, The Queen's University, Belfast BT7 1NN, United Kingdom \\
$^4$Center for Subwavelength Optics and Department of Physics and Astronomy, Seoul National
University, Seoul, 151-742, South Korea \\}

\date{\today}

\begin{abstract}
We present the experimental realization of a scheme, based on single-photon interference, for
implementing superpositions of distinct quantum operations. Its application to a thermal light field (a
well-categorized classical entity) illustrates quantum superposition from a new standpoint and provides
a direct and quantitative verification of the bosonic commutation relation between creation and
annihilation operators. By shifting the focus towards operator superpositions, this result opens
interesting alternative perspectives for manipulating quantum states.
\end{abstract}


\maketitle

The superposition principle is one of the pillars upon which the entire structure of quantum mechanics
is built~\cite{dirac}. A quantum system in a pure state can always be described as a superposition of
linearly independent states; thus once one has a quantum system represented by a pure state, the
superposition is naturally there. An inter-body superposition state, the so-called entangled state, is
somewhat trickier to generate than a single-body superposition state. However, it has been demonstrated
that entanglement can be achieved by various methods, including a series of unitary
operations~\cite{sackett00,haeffner08,yamamoto03} or by post-selection of events after unitary
operations~\cite{pan08}. On the other hand, the discussion about superpositions of classical mixed
states is not as clear as for a pure state~\cite{jeong06}.

Quantum operators, besides quantum states, play a crucial role in describing physical operations
including unitary transformations and measurements in quantum theory. If one can implement a
superposition of operators, one can also construct state superpositions by applying the superposed
operators to a given state, unless it is a simultaneous eigenstate of the component operations. In fact,
also the Schr\"{o}dinger's cat paradox~\cite{schroedinger35} can be understood as the quantum-mechanical
impact of the superposition of macroscopically distinct operations (\textit{to kill} or \textit{not to
kill}) on a classical object (the \textit{cat}).

Several groups have recently succeeded in applying simple quantum operators to different quantum states.
For example, in the optical domain, basic operations, like single-photon addition and subtraction, have
been demonstrated to produce highly nonclassical~\cite{agarwal91,zavatta04,zavatta05,ourjoumtsev06} and
non-Gaussian states~\cite{wenger04} even when applied to classical states of
light~\cite{agarwal92,zavatta07}. Both photon addition and subtraction are performed in a conditional
way upon the detection of a single photon in an ancillary (herald) light mode. Sequences of photon
additions and subtractions have also been implemented to show that the two sequences $\hat a \hat
a^{\dag}$ and $\hat a^{\dag} \hat a$, where $\hat a^{\dag}$ and $\hat a$ are the bosonic creation and
annihilation operators, give different results when applied to the same input light
state~\cite{parigi07}. This is an important corner stone for the proof of the bosonic commutation
relation
\begin{equation}
[\hat a,\hat a^{\dag}]=\hat a \hat a^{\dag} - \hat a^{\dag} \hat a= \openone \label{comm}
\end{equation}
which is at the heart of many important consequences of quantum mechanics. However, the complete
demonstration of the commutation relation was out of reach because of the lack of an important element
in the quantum manipulation toolbox: the possibility of superposing different operators $\hat A$ and
$\hat B$ to form the general operator superposition $\alpha \hat A + \beta \hat B$, where $\alpha$ and
$\beta$ are complex amplitudes.

Since the superposition principle relies on the indistinguishability among different alternatives, the
experimental implementation of quantum operators heralded by a single-photon detection offers a very
convenient way to achieve this goal. If the herald field modes of different operators are properly mixed
by means of a beam splitter, the information about the origin of a click in the herald photodetector is
erased and a coherent superposition of the different operators can be conditionally implemented.
Somewhat similar schemes have been recently proposed and experimentally implemented for increasing the
entanglement of bipartite Gaussian quantum states by inconclusive photon
subtraction~\cite{olivares03,ourjoumtsev07,ourjoumtsev09}, toward the implementation of a quantum
repeater for long-haul quantum communication~\cite{kimble08} in ionic systems~\cite{yuan08}, and for the
remote delocalization of a single photon over distinct temporal modes~\cite{zavatta06}.

In this Letter we present the experimental realization of a general scheme, based on single-photon
interference, for superposing distinct quantum operations. As recently proposed in Ref.~\cite{kim08},
demonstrating the bosonic commutation relation thus reduces to realizing the balanced superposition
$\hat a \hat a^{\dag} - \hat a^{\dag} \hat a$, and showing that it corresponds to the identity operator
$\openone$. While most of the mathematical structure of quantum mechanics is based on the commutation
relation, this is the first time it is directly probed in an experiment.

The primary laser source is a mode-locked Ti:Sa laser emitting 1.5 ps pulses at a repetition rate of 82
MHz. A rotating ground glass disk (RD) is inserted in the path of the laser beam and a bare single-mode
fibre (SMF) is used to collect a portion of the scattered light to provide the pulsed thermal light
states~\cite{parigi09} of mean photon numbers around unity which have been used as the input states in
the experiment. We use a convenient modular scheme (see Figure 1), where one single-photon addition
stage~\cite{zavatta05} ($\hat a^{\dag}$, based on conditional single-photon parametric amplification in
a type-I BBO -$\beta$ barium borate- nonlinear crystal) is placed between two single-photon subtraction
stages~\cite{wenger04} ($\hat a$, based on the conditional removal of a single photon by adjustable
low-reflectivity beam splitters, obtained with combinations of half-wave plates (HWP) and polarizing
beam-splitters (PBS)).
\begin{figure}[h]
\includegraphics [width=8.5cm]{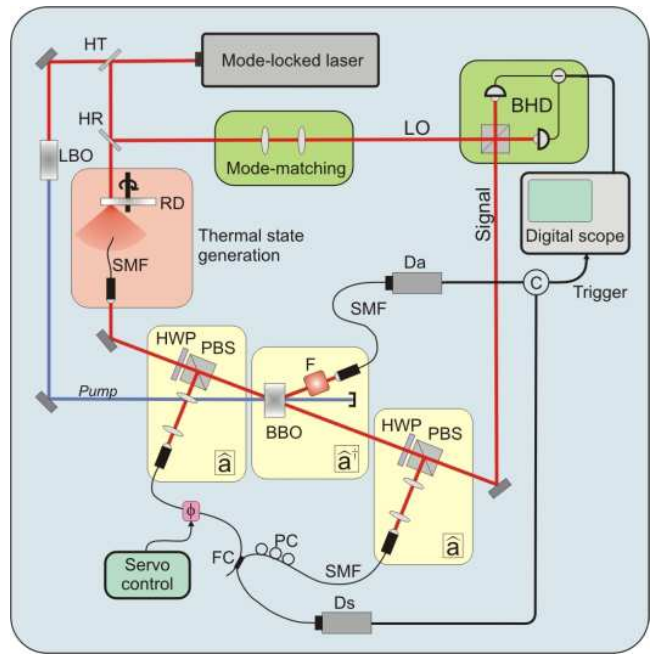}
\caption{Experimental setup. Pump pulses for parametric down-conversion are obtained by
frequency-doubling the laser output in a LBO -lithium triborate- crystal. The conditionally-prepared
signal state is mixed with a strong reference coherent field (LO, obtained from a portion of the main
laser output) on a 50-50 beam splitter whose outputs are detected by two photodiodes (Hamamatsu PIN
S3883). The balanced homodyne detection (BHD) signal is acquired and stored by a digital oscilloscope
(Tektronix TDS7104) on a pulse-to-pulse basis triggered by a coincidence (C) between clicks from the Da
and Ds single-photon detectors (Perkin Elmer model SPCM AQR-14). } \label{exp}
\end{figure}

The successful implementation of the desired superposition of operator sequences is determined by
the coincidence (C) between a click from the single-photon detector heralding photon addition (Da)
and one from a second photodetector (Ds), placed after a -3db fibre coupler (FC) combining the
herald signals from the two subtraction modules. By a click from detector Ds we know that a single
photon has been subtracted, but we cannot identify if it was before or after the photon addition.
In these conditions, a coincidence event heralds the application of the general operator
superposition $\hat{a} \hat{a}^{\dag} -e^{i \phi}\hat{a}^{\dag} \hat{a}$ with an adjustable phase
$\phi$, to any input light field. By varying the phase $\phi$ with a piezo-actuated mirror, any
arbitrary balanced superposition of the two operator sequences can be obtained. In particular, by
setting $\phi=0$ or $\phi=\pi$, one can directly implement the commutator or the anti-commutator of
the creation and annihilation operators, respectively. Note that the present scheme, differently
from the original theoretical proposal~\cite{kim08}, where only these two possibilities were
allowed by using two detectors at both exits of the beam splitter, allows for greater flexibility
(by generating operator superpositions with arbitrary relative phases) and experimental simplicity.

Without the click from the addition module ($\hat a^{\dag}$), the scheme reduces to a
Mach-Zehnder-type interferometer which can be used to verify the indistinguishability of the two
subtraction events by evaluating the visibility of the single-photon interference in the counts at
detector Ds. Visibilities of about $97\%$ are obtained by carefully balancing the reflectivity
($\approx 3 \%$) of the involved beam splitters, by fine polarization control (PC), and by a
precise adjustment of the delays between the corresponding herald modes. If delays are not
compensated, only a statistical mixture of the two operator sequences with equal weights is
obtained. The Ds count rate is also used to monitor the superposition phase $\phi$ and lock it to
any desired value. The effects of experimental deviations from the ideal realization of this scheme
(such as the finite reflectivity of the subtraction beam splitters, the possible multiple
photon-pair production in the parametric process, or the fact that real photodetectors are not able
to discern the number of photons but only there being photons or not) have already been
shown~\cite{zavatta05,kim08:jpb} not to significantly affect the results of the experiment for the
present range of parameters.

The state resulting from the chosen operator superposition is analyzed by means of a
high-frequency, time-domain, balanced homodyne detector~\cite{josab02} yielding the distributions
of measured field quadratures. Since both the initial thermal states and those resulting from the
above manipulations possess no intrinsic phase, the phase of the local oscillator (LO, the
reference coherent field for homodyne detection) is not actively scanned, and phase-independent
marginal distributions are obtained. However, the final states still clearly depend on the phase
$\phi$ of the superposition. High experimental efficiency is obtained by minimizing all spurious
losses and making sure that all the single-photon operations are performed in exactly the same
spatiotemporal mode as the one selected by the LO. This requires narrow spatial and spectral
filtering (F) in the herald mode of the parametric down-conversion crystal, and an accurate
matching of the fibre-coupled fields reflected from the two subtracting beam splitters to the LO
spatial mode.

Figure 2a) shows a sequence of histograms of raw homodyne data acquired while scanning the phase of
the superposition. The remote manipulation of the state by the implementation of different
superpositions of creation and annihilation sequences is clearly observed. The quadrature
distribution of the final state undergoes a very rapid initial evolution from a bell-shaped curve
at $\phi=0$ towards a volcano-shaped one around $\phi=\pi$, where the phase dependence is much
slower. The phase change would not have resulted in such different output states if the operations
had been statistical mixtures.
\begin{figure}[h]
\includegraphics [width=8cm]{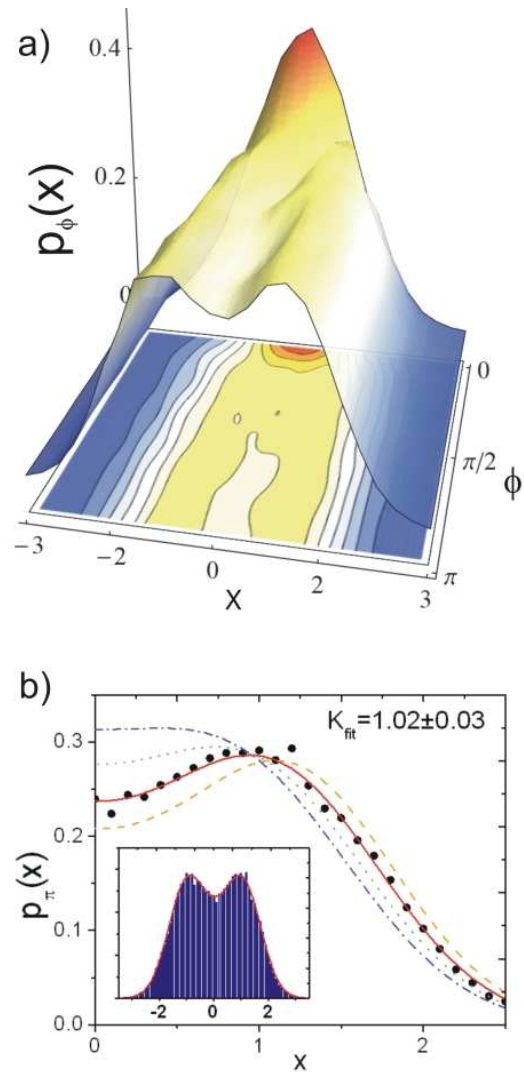}
\caption{a) Sequence of histograms corresponding to about 250,000 raw quadrature data for the final
state as a function of the normalized quadrature x and of the superposition phase $\phi$. Data have been
obtained in a 16-hour measurement, and have been binned in 9 phase intervals between 0 and $\pi$. A
thermal field with a mean photon number $\bar n =0.9(1)$ is used as the initial state.  b) Histogram of
raw quadrature data (solid dots) for the anti-commutator setup at $\phi=\pi$. Also shown are theoretical
curves calculated for the actual experimental parameters (total efficiency $\eta=0.61$, $\bar n =0.9$)
and different values of the commutator ($K=0$: dashed orange; $K=2$: dotted green; $K=3$: dash-dotted
blue). The solid red curve is the result of the best fit to the experimental data.} \label{data}
\end{figure}

The special cases of $\phi=0$ and $\phi=\pi$ are illustrated in more detail in Figure 3, where the
Wigner functions of the original thermal state and those resulting from the experimental realization of
the commutator and anti-commutator between $\hat a$ and $\hat a^{\dag}$ are presented.
\begin{figure}[h]
\includegraphics [width=8.5cm]{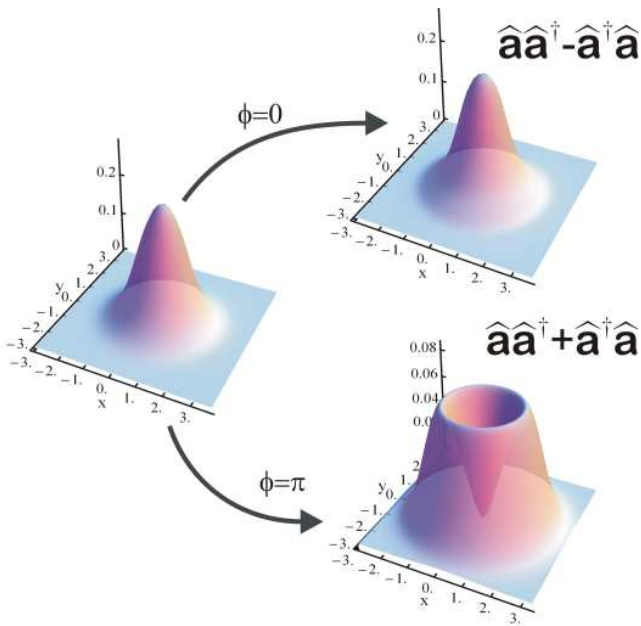}
\caption{Experimentally reconstructed Wigner functions of the original thermal state and of those
resulting from the application of the commutator and anti-commutator superpositions. The state is not
changed for $\phi=0$. About $10^4$ ($10^5$) quadrature data points have been acquired in the
(anti-)commutator case.} \label{wig}
\end{figure}
The fidelity $F=|Tr \sqrt{\sqrt{\hat \rho_{in}} \hat \rho_{out} \sqrt{\hat \rho_{in}}}|^2$ between
the original thermal state and the final one (represented by their reconstructed density operators
$\rho_{in}$ and $\rho_{out}$, respectively) is about $F=0.992$ for the commutator case ($\phi=0$).
This demonstrates that the implemented operator superposition is essentially equivalent to the
identity operator. Wigner functions have been obtained from the 10 diagonal density matrix elements
(13 for the anti-commutator case) reconstructed by means of a maximum likelihood
algorithm~\cite{lvovsky04,hradil06} without any correction for the finite detection efficiency. If
homodyne detection efficiency ($\eta_d=0.7$) is corrected for, the Wigner function for the state
resulting from the anti-commutation operator ($\phi=\pi$) clearly attains negative values. The
existence of negative regions in the reconstructed Wigner function is a direct signature of the
fact that the state impinging on the homodyne detector is highly nonclassical.

Actually, it is interesting to note a fact that was not realized in the theoretical
proposal~\cite{kim08}: because of the normalization of quantum states, the above results just
demonstrate the commutation relation up to a multiplicative constant $K$, i.e., one might still have
$[\hat a,\hat a^{\dag}]=K \openone$. However, in this case, the anti-commutator setup implements the
$2\hat a^{\dag} \hat a + K \openone$ operator, which produces an output state strongly depending on the
exact value of the constant $K$. Figure 2b) reports the measured homodyne quadrature distribution for
the same initial thermal state after the application of the anti-commutation operator ($\phi=\pi$). Also
reported are the theoretical distributions calculated for the same experimental parameters but with a
few different values of the constant $K$. Experimental data are consistent with $K=1$, whereas different
integer values are in evident disagreement. A best fit of the experimental homodyne data gives
$K=1.02(3)$, thus quantitatively demonstrating the bosonic commutation relation.

Although the present case only required a coherent superposition of two (sequences of) quantum
operators with the same weight, the proposed scheme is much more general and allows one, in
principle, to implement coherent superpositions of an arbitrary number of operators with arbitrary
relative amplitudes and phases by a network of beam splitters with adjustable reflectivities. The
single-photon interference as a way to produce general operator superpositions can be
straightforwardly extended to the superposition of $\hat a$ and $\hat a^{\dag}$ by letting the
creation and annihilation herald photons interfere at a beam splitter. As a representative example,
"position" and "momentum" operators in the phase space can be implemented in this way. Our approach
can even be generalized to realize various superpositions of higher-order operators in terms of
$\hat a$ and $\hat a^{\dag}$. Any quantum state can be written as $\sum_n C_n \hat{a}^{\dag n}
\ket{0}$, where $C_n$ are complex amplitudes, and arbitrary states can thus be generated by
applying the appropriate superposition of photon creation operators. In this paper we have
experimentally demonstrated a basic building unit for such general operator superpositions on a
traveling light field for the first time.

{\it Acknowledgements}-Work performed in the frame of the "Spettroscopia laser e ottica
quantistica" project of the Department of Physics of the University of Florence, with the support
of Ente Cassa di Risparmio di Firenze and CNR-RSTL. M.S.K. acknowledges UK EPSRC and QIP IRC at
Oxford for financial support. H.J. was supported by the WCU program and the KOSEF grant funded by
the Korea government (MEST)(R11-2008-095-01000-0).

\bibliography{Fock_bib}
\end{document}